\documentclass[12pt]{article}
\pdfoutput=1
\usepackage{graphicx}
\usepackage{amsmath, amssymb}
\usepackage{slashed}
\usepackage{xcolor}
\usepackage{array,arydshln}
\usepackage{cite}
\usepackage{enumerate}
\usepackage[italicdiff]{physics}
\usepackage[pdftex,
  colorlinks=true,
citecolor=green!60!blue]{hyperref}
  
\begin{document}

\begin{titlepage}

\begin{center}

\hfill KANAZAWA-23-04 \\
\hfill March 2023

\vspace{0.5cm}

{\Large\bf Supersymmetric Baryogenesis in a \\
  Hybrid Inflation Model}

\vspace{1cm}
{\large Yoshihiro Gunji,}$^{\it (a)}$ 
{\large Koji Ishiwata,}$^{\it (a)}$
{\large Takahiro Yoshida}$^{\it (b)}$

\vspace{1cm}

{\it $^{(a)}$Institute for Theoretical Physics, Kanazawa University, Kanazawa
  920-1192, Japan}

{\it $^{(b)}$Department of Information, Kaishi Professional University, Niigata 950-0916, Japan}

\vspace{1cm}

\abstract{We study baryogenesis in a hybrid inflation model which is
  embedded to the minimal supersymmetric model with right-handed
  neutrinos. Inflation is induced by a linear combination of the
  right-handed sneutrinos and its decay reheats the universe. The
  decay products are stored in conserved numbers, which are
  transported under the interactions in equilibrium as the temperature
  drops down. We find that at least a few percent of the initial
  lepton asymmetry is left under the strong wash-out due to the
  lighter right-handed (s)neutrinos. To account for the observed
  baryon number and the active neutrino masses after a successful
  inflation, the inflaton mass and the Majorana mass scale should be
  $10^{13}\,{\rm GeV}$ and
  $\order{10^{9}\mathchar`-\mathchar`-10^{10}}\,{\rm GeV}$,
  respectively. }

\end{center}
\end{titlepage}

\renewcommand{\theequation}{\thesection.\arabic{equation}}
\renewcommand{\thepage}{\arabic{page}}
\setcounter{page}{1}
\renewcommand{\thefootnote}{\#\arabic{footnote}}
\setcounter{footnote}{0}

\section{Introduction}
\label{sec:intro}
\setcounter{equation}{0}

The baryon number of the universe is precisely determined by the
observations. For example, the latest result by the Planck
collaboration gives~\cite{Aghanim:2018eyx}
\begin{align}
  (n_B/s)_{\rm obs}=
  (8.70\pm0.04)\times 10^{-11}\,,
  \label{eq:eta_obs}
\end{align}
where $n_B$ and $s$ are the number density of the baryon and the
entropy density of the present universe, respectively.  The genesis of
the baryon number is one of the mysteries of our universe since the
standard model of particle physics can not explain the observed value
of the baryon number. Leptogenesis~\cite{Fukugita:1986hr} is a viable
mechanism to generate the baryon number. The heavy right-handed
neutrinos are introduced in addition to the standard model particles
and their decay and scattering in the thermal plasma produce the
sufficient lepton density.  Finally the lepton number is converted to
the baryon number by the sphaleron process. This mechanism is
economical in a sense that the heavy right-handed neutrinos also
explain the tiny neutrino masses
naturally~\cite{Minkowski:1977sc,Yanagida:1979as,Yanagida:1980xy,Gell-Mann:1979vob,Ramond:1979py,Glashow:1979nm},
called seesaw mechanism.

In this work, we study baryogenesis in a supersymmetric model
motivated by the inflation. This model is based on the superconformal
subcritical hybrid inflation
model~\cite{Ishiwata:2018dxg,Gunji:2021zit}, where inflation continues
even after the inflaton field becomes below the critical point value
of the hybrid inflation. Such subcritical regime of inflation is
originally considered in
Refs.~\cite{Buchmuller:2014rfa,Buchmuller:2014dda} with an approximate
shift symmetry in K\"ahler potential.\footnote{In the scenario, the
waterfall field value is suppressed and inflation continues in the
infaton direction, meanwhile
Refs.\,\cite{Clesse:2010iz,Clesse:2012dw,Kodama:2011vs} study the case
where inflation along the direction of the waterfall field. }
Refs.\,\cite{Ishiwata:2018dxg,Gunji:2021zit} consider the
superconformal model combined with the approximate shift symmetry and
found the model gives a good fit with the observed spectral index
of the scalar amplitude and the tensor-to-scalar ratio, and that the
inflaton mass is predicted to be around $10^{13}\,{\rm GeV}$. In the
current study, we embed the inflation model to the minimal
supersymmetric standard model with right-handed neutrinos. The model
is the same as one considered in Ref.\,\cite{Gunji:2019wtk} and we
extend the study to a more realistic scenario by taking into account
the flavor
effects~\cite{Barbieri:1999ma,Abada:2006fw,Nardi:2006fx,Abada:2006ea,Blanchet:2006be,Antusch:2006cw}
and the spectator
effects~\cite{Buchmuller:2001sr,Nardi:2005hs,Garbrecht:2014kda}.
After inflation, the inflaton decays to reheat the universe and
produce the lepton numbers. The important points are {\it i)} not all
the lepton numbers are washed out due to the lighter right-handed
(s)neutrinos~\cite{Engelhard:2006yg,Bertuzzo:2010et} and {\it ii)} the
$B-L$, where $B$ and $L$ are baryon and lepton number respectively,
remains due to the conserved charges even in a case where the wash-out
effect is most effective~\cite{Domcke:2020quw,Fong:2020fwk}. In the
latter point the supersymmetry plays the crucial role.

\section{The model}
\label{sec:model}
\setcounter{equation}{0} 

We consider an extended minimal supersymmetric standard model (MSSM)
augmented by three right-handed neutrinos $N_i$ and two standard model
singlet fields $S_\pm$, which are described by a superpotential
\begin{align}
   W \supset
  \frac{1}{2}M_{ij}N_iN_j+y_{ij}N_i L_j H_u
  +\lambda_i N_i S_+ S_-\,,
  \label{eq:W_0}
\end{align}
where $y_{ij}$ and $\lambda_i$ are coupling constants and $M_{ij}$ are
the Majorana masses. The indices $i,j$ take $1,2,3$. $L_i$ and
$H_u$ are the left-handed lepton doublets and the up-type Higgs,
respectively.  In this model we assume that $S_\pm$ has a local U(1)
charge, $\pm q$, and the other fields are the U(1) singlets, and that
the gauge symmetry is spontaneously broken due to the D-term
potential, which is studied in Ref.\,\cite{Gunji:2019wtk}.
Consequently it acquires nonzero vacuum expectation value (VEV),
denoted as $\expval{S_+}$. Then in $(N_i, S_-)^T$ basis, we have the
following 4 by 4 mass matrix,
\begin{align}
  M_N&\equiv
\left(
\begin{array}{cccc}
&&&\lambda_1\langle S_+\rangle \\
  &\mbox{\smash{\Large $M$}}&&\lambda_2\expval{S_+}\\
&&& \lambda_3\expval{S_+}\\ 
  \lambda_1\langle S_+\rangle &\lambda_2\expval{S_+}
  &\lambda_3\expval{S_+}  & 0
\end{array}
\right)\,,
\end{align}
which leads to a 3 by 3 active neutrino mass matrix,
\begin{align}
  M_\nu&=-\expval{H_u}^2\tilde{y}^TM_N^{-1}\tilde{y},
  \\
\tilde{y}&=
      \left(
\begin{array}{ccc}
&& \\
  &\mbox{\smash{\Large $y$}}&\\
&& \\ 
 0&0&0
\end{array}
\right)\,.
\end{align}
This matrix is diagonalized by a unitary matrix $U_\nu$ as
\begin{align}
  U^T_\nu M_\nu U_\nu ={\rm diag}(m_1,m_2,m_3)\,.
  \label{eq:M_nu}
\end{align}
It is known that the matrix $M_\nu$ gives rise to one massless
neutrino~\cite{Gunji:2019wtk}. We follow the convention such that
$m_1=0$ and $m_3>m_2$ for the normal hierarchy (NH) and $m_3=0$ and
$m_2>m_1$ for the inverted hierarchy (IH). Another important fact is
that $M_\nu$ is independent of both the mass scale $\lambda_i
\expval{S_+}$ and $y_{3i}$. Therefore, $\lambda_i \expval{S_+}$ and
$y_{3i}$ are not constrained by the observed neutrino masses. This is
crucial in the later discussion.

For later convenience, we introduce two bases; inflaton basis and mass
eigenstate basis. The former one, written as $(N'_1, N'_2, N'_3)^T$,
is a basis where $N'_3$ only couples to $S_+S_-$. Namely
$\lambda_iN_i\equiv \tilde{\lambda}N_3'$ and the scalar component of
$N'_3$ plays the role of the inflaton field.  In $(N'_i, S_-)^T$
basis, we have the following 4 by 4 mass matrix,
\begin{align}
  M'_N&=
      \left(
\begin{array}{cccc}
&&& 0 \\
  &\mbox{\smash{\large $U_{\rm inf}^*MU_{\rm inf}^\dagger$}}&& 0\\
&&& m_\phi\\ 
  0&0 &m_\phi  & 0
\end{array}
\right)\,, 
\end{align}
where $N'_i=U_{{\rm inf}\,ij}N_j$ and
$m_\phi=\tilde{\lambda}\expval{S_+}$, which corresponds to the
inflaton mass.  In this basis, $\tilde{y}$ transforms as
\begin{align}
  \tilde{y}'&=
      \left(
\begin{array}{ccc}
&& \\
  &\mbox{\smash{\Large $U_{\rm inf}^*y$}}&\\
&& \\ 
 0&0&0
\end{array}
\right)\,.
\end{align}

The latter one is the basis where both $M'_N$ and
charged lepton mass matrix are diagonalized. In the basis, the relevant
terms in the superpotential are
\begin{align}
  W \supset \frac{1}{2}M_I\hat{N}_I\hat{N}_I+\hat{y}_{Ij}\hat{N}_I
  \hat{L}_jH_u\,,
\end{align}
where $\hat{N}_I$ ($I=1,2,3,4$) and $\hat{L}_i$ are the mass
eigenstates of the heavy right-handed neutrinos plus $S_-$ and the
charged leptons.\footnote{We will sometimes use a notation
$\hat{L}_\alpha$ where $\alpha=e,\mu,\tau$ in the later discussion.}
Namely,
\begin{align}
  &\hat{N}=U_N^\dagger N'\,,~~~\hat{L} = T^\dagger L\,,\\
  &\hat{y}=
  U^T_N\left(
\begin{array}{ccc}
&& \\
  &\mbox{\smash{\large $U^*_{\rm inf}y$}}&\\
&& \\ 
 0&0&0
\end{array}
\right)T\,.
\end{align}
Here we have defined a unitary matrix $U_N$ as
\begin{align}
 U^T_N M'_N U_N={\rm diag}(M_1,M_2,M_3,M_4)\,,
\end{align}
where $M_1<M_2<M_3<M_4$ and similar for $T$. Then the PMNS matrix is
given by $U_{\rm PMNS}=T^\dagger U_{\nu}$.

In order not to disturb the inflationary trajectory,  we assume
\begin{align}
  \lambda_i \expval{S_+}\gg \Lambda \,,~~ M_{ij}\sim \order{\Lambda}\,,
  \label{eq:mphi>>M}
\end{align}
where $\Lambda$ represents the typical scale of the Majorana
masses. In this limit, the mass eigenvalues $M_I$ have a relation
\begin{align}
  M_1\sim M_2\sim \Lambda \,,~~m_\phi\simeq M_3\simeq M_4\,,
  \label{eq:M3=M4}
\end{align}
and $U_N$ has a structure as
\begin{align}
  U_N= \left(
  \begin{array}{c:c}
    u_{2\times 2} &\order{\frac{\Lambda}{m_\phi}} \\
    \hdashline
    \order{\frac{\Lambda}{m_\phi}}    & 
        {\scriptsize
          \frac{1}{\sqrt{2}}
          \mqty(i & 1\\
            -i & 1)+\order{\frac{\Lambda}{m_\phi}}
            }
  \end{array}
  \right)\,,
  \label{eq:UN_app}
\end{align}
where $u_{2\times 2}$ is $\order{1}$ 2 by 2
matrix. Eq.\,\eqref{eq:mphi>>M} is the feature of this model and
Eqs.\,\eqref{eq:mphi>>M}--\eqref{eq:UN_app} are important in the
estimation of the lepton asymmetry. As we described below
Eq.\,\eqref{eq:M_nu}, the active neutrino mass matrix is independent
of $m_\phi$. Thus $m_\phi$ is not constrained by the observation of
the neutrino masses. We will see in the next section that
$\hat{y}_{3i}$ are free from the constraint, which means that
$\hat{y}_{3i}$ are free parameters in this model.

\section{Baryogenesis}
\label{sec:baryogenesis}
\setcounter{equation}{0}

The overview of the thermal history of our model is the following:
\begin{enumerate}[{\it a)}]
\item scalar component of $N'_3$, denoted as $\tilde{N}'_3$, drives
  inflation
\item the inflaton decays to reheat the universe and produce lepton
  number non-thermally
\item part of the lepton number is washed out by $\hat{N}_1$ and
  $\hat{N}_2$\footnote{The decays of $\hat{N_1}$ and $\hat{N_2}$ might
give comparable contributions by a tuning of the model
parameters~\cite{Giudice:2003jh}. To be conservative we ignore them. }
\item the lepton number is converted to baryon number by the sphaleron process
\end{enumerate}
At the stage {\it a)} we can take $\sqrt{2}\Re\,\tilde{N}'_3=\phi$ as
the inflaton without the loss of generality and $\phi$ drives
inflation.  Such inflation models are discussed in
Refs.\,\cite{Gunji:2019wtk,Gunji:2022xig}. After inflation, the
inflaton field decays to reheat the universe.  The lepton number is
produced simultaneously by the inflaton decay, which is given by
\begin{align}
  L^{\rm dec}\equiv
  \frac{n_L}{s}\Bigl |_{\rm dec}
  = \frac{3}{4}\frac{T_R}{m_\phi}\epsilon_\phi\,,
  \label{eq:YLdec}
\end{align}
where $T_R$ and $\epsilon_\phi$ are the reheating temperature and the
lepton asymmetry of the inflaton decay, respectively.\footnote{This is
similar to right-handed sneutrino inflation and
leptogenesis~\cite{Murayama:1992ua,Murayama:1993xu,Murayama:1993em,Hamaguchi:2001gw,Ellis:2003sq,Antusch:2004hd,Antusch:2009ty,Kadota:2005mt,Nakayama:2013nya}. As
we will see, however, the thermal history in our model is different
from those discussed in the literature.} This corresponds to the stage
{\it b)}.  Assuming the instantaneous reheating and
$T_R/m_\phi\lesssim 1$, the reheating temperature is given by the
decay width $\Gamma_\phi$ of the inflaton as $T_R\simeq
(90/\pi^2g_*(T_R))^{1/4}\sqrt{\Gamma_\phi M_{\rm Pl}}$ where
$g_*(T_R)\simeq 228.75$ and
\begin{align}
  \Gamma_\phi\simeq \frac{(\hat{y}\hat{y}^\dagger)_{33}}{4\pi} m_\phi\,,
\end{align}
where we have used $\Lambda/m_\phi\ll 1$, given in
Eq.\eqref{eq:mphi>>M}. Similarly, the asymmetry $\epsilon_\phi$ is
given by
\begin{align}
  \epsilon_\phi&\simeq \frac{1}{8\pi}\sum_{K\neq 3}
  \frac{\Im [\{(\hat{y}\hat{y}^\dagger)_{3K}\}^2]}
       {(\hat{y}\hat{y}^\dagger)_{33}}g(x_K)
       \nonumber \\
       &\simeq -\frac{1}{8\pi}
       \frac{\Im [\{(\hat{y}\hat{y}^\dagger)_{34}\}^2]}
       {(\hat{y}\hat{y}^\dagger)_{33}}\frac{m_\phi}{\Delta M}\,,      
\end{align}
where $g(x)=(\frac{2}{1-x}-\ln \frac{1+x}{x})\sqrt{x}$,
$x_K=M_K^2/M_3^2$ and $\Delta M=M_4-M_3\sim \order{\Lambda}$.  The
second line comes from $x_{1,2}\ll 1$ and $x_4\simeq 1$. Then
$g(x_4)\simeq -M_3/\Delta M\simeq -m_\phi/\Delta M$ gives the dominant
contribution.  Here we have used the fact
\begin{align}
  \frac{(\Gamma_\phi M_3)^2}{(M_3^2-M_4^2)^2}
  \simeq \left[\frac{\Gamma_\phi}{2\Delta M}\right]^2
  \sim \left[4\times 10^{-5}
    ~~
    \frac{(\hat{y}\hat{y}^\dagger)_{33}}{10^{-6}}
    \frac{m_\phi}{10^{13}\,{\rm GeV}}
    \frac{10^{10}\,{\rm GeV}}{\Lambda}
    \right]^2\ll 1\,,
\end{align}
in the parameter space we are interested in. 

To estimate $T_R$ and
$\epsilon_\phi$, it is convenient to introduce a 4 by 3 matrix $R$ based on
Ref.\,\cite{Casas:2001sr}:
\begin{align}
  R=iD^{-1/2}_N U_{N}^T \tilde{y}'\expval{H_u} U_\nu D_{\nu}^{-1/2}\,,
\end{align}
where $D^{\pm 1/2}_N={\rm diag}(M^{\pm 1/2}_1,M^{\pm 1/2}_2,M^{\pm
  1/2}_3,M^{\pm 1/2}_4)$, and $D_\nu^{\pm 1/2}={\rm diag}(0,m^{\pm
  1/2}_2,m^{\pm 1/2}_3)$ for the NH, ${\rm diag}(m^{\pm 1/2}_1,m^{\pm
  1/2}_2,0)$ for the IH. $R$ satisfies $R^TR={\rm diag}(0,1,1)$ and
${\rm diag}(1,1,0)$ for the NH and IH, respectively.  From the
equation, we write $\hat{y}$ in terms of $R$ as
\begin{align}
  \hat{y}\expval{H_u}=-i D^{1/2}_N R D^{1/2}_\nu U^\dagger_{\rm PMNS}\,,
\end{align}
which leads to
\begin{align}
  (\hat{y}\hat{y}^\dagger) \expval{H_u}^2 =
  D_N^{1/2} R D_\nu R^\dagger D_N^{1/2}\,.
  \label{eq:yydg}
\end{align}
In the expansion of $\Lambda/m_\phi$, we find
\begin{align}
  (\hat{y}\hat{y}^\dagger)_{33}  &=
  m_\phi\sum_{i} m_i |R_{3i}|^2/\expval{H_u}^2\,,
  \\
  (\hat{y}\hat{y}^\dagger)_{34}  &=
  i(\hat{y}\hat{y}^\dagger)_{33}(1 + \order{\Lambda/m_\phi})\,,
  \label{eq:yydg_34}
\end{align}
where we have used Eq.\,\eqref{eq:UN_app}. It is clear that
$\hat{y}_{3i}$ is not constrained by the observed results in the
neutrino sector. We note that imaginary part of
$\{(\hat{y}\hat{y}^\dagger)_{34}\}^2$ is suppressed by
$\Lambda/m_\phi$ compared with the naive expectation, which cancel a
factor of $m_\phi/\Delta M$.
Thus, we get
\begin{align}
  T_R&\simeq 6\times 10^{11}\,{\rm GeV}
  \sqrt{\frac{(\hat{y}\hat{y}^\dagger)_{33}}{10^{-6}}
    \frac{m_\phi}{10^{13}\,{\rm GeV}} }\,,
  \label{eq:TR} \\
  \epsilon_\phi &\simeq
  \frac{a}{4\pi}(\hat{y}\hat{y}^\dagger)_{33}
  \simeq 8\times 10^{-8}a\frac{(\hat{y}\hat{y}^\dagger)_{33}}{10^{-6}}\,.
  \label{eq:epsilon_phi}
\end{align}
Here we have introduced a coefficient $a$ to take into account
$\order{\Lambda/m_\phi}$ term in Eq.\,\eqref{eq:yydg_34}. We expect
$a=\order{1}$ without a fine-tuning. $T_R/m_\phi$ can be written in
terms of $K_\phi$ as $T_R/m_\phi\simeq \sqrt{K_\phi}$.  Here
$K_\phi\equiv K_3\simeq K_4$, where $K_I\equiv \tilde{m}_I/m_*$ and
\begin{align}
  \tilde{m}_I\equiv
  \frac{(\hat{y}\hat{y}^\dagger)_{II}\expval{H_u}^2}{M_I}\,,~~~
  m_*=\frac{4\pi^2\sqrt{g_*(M_I)}\expval{H_u}^2}{3\sqrt{10}M_{\rm Pl}}\,.
\end{align}
Thus the condition $T_R/m_\phi\lesssim 1$ is equivalent to
$K_\phi\lesssim 1$.

At the stage {\it c)}, the generated lepton number suffers from the
wash-out by $\hat{N}_1$ and $\hat{N}_2$. To evaluate the wash-out
effect, we estimate $K_{1,2}$. From Eq.\,\eqref{eq:yydg}, it is
straightforward to obtain 
\begin{align}
  \tilde{m}_I =\sum_i m_i |R_{Ii}|^2 \ge m_{\rm min}\,,
\end{align}
for $I=1,2$ where $m_{\rm min}=m_2\simeq 8.6\times 10^{-3}\,{\rm eV}$
and $m_1\simeq 4.9\times 10^{-2}\,{\rm eV}$ for the NH and IH cases,
respectively~\cite{Esteban:2020cvm}. Here we have taken
$\Lambda/m_\phi\ll 1$ to derive $m_{\rm min}$. Therefore, $K_{1,2}$
have the minimum values as
\begin{align}
  K_I \ge \left\{
  \mqty{22 & ({\rm NH}) \\
  124 & ({\rm IH})}
  \right.~~~~~~ {\rm for~}I=1,2\,.
\end{align}
This means that the wash-out effect is strong. Even in the strong wash-out
regime, not all lepton number is washed
out~\cite{Engelhard:2006yg,Bertuzzo:2010et}.  At the production of the
lepton number we assume
\begin{align}
  m_\phi\gtrsim 10^{13}\,{\rm GeV}\,.
\end{align}
This requirement is for a successful inflation, which will be
quantified later in Eq.\,\eqref{eq:inflaton_mass_range}. The mass
scale of inflaton means that the all Yukawa interactions are out of
equilibrium at the decay of the inflaton, except for the top Yukawa
interaction. Consequently the produced lepton is a coherent state
$\ket*{\ell_3}(\simeq \ket*{\ell_4})\equiv \ket*{\ell_{\phi}}$,
defined by
$\ket*{\ell_I}=\frac{1}{\sqrt{(\hat{y}\hat{y}^\dagger)_{II}}}
\hat{y}_{I\alpha}\ket*{\ell_\alpha}$\,.  As the temperature drops
down, the spectator
effects~\cite{Buchmuller:2001sr,Nardi:2005hs,Garbrecht:2014kda}, the
flavor
effects~\cite{Barbieri:1999ma,Abada:2006fw,Nardi:2006fx,Abada:2006ea,Blanchet:2006be,Antusch:2006cw}
and the wash-out effect due to $\hat{N}_1$ and $\hat{N}_2$ become
important.

Regarding the masses of the lighter right-handed neutrinos, we
consider
\begin{align}
  10^{7}\,{\rm GeV} \lesssim M_{1,2} \lesssim 10^{10}\,{\rm GeV}\,.
  \label{eq:M_{1,2}}
\end{align}
Here the upper bound is from the requirement~\eqref{eq:mphi>>M},
meanwhile the lower one is to ignore the $\mu$ term, i.e., $\mu
H_uH_d$, and the gaugino masses. To be more quantitative, $M_{1,2}$
should be larger than roughly $2\times 10^{7}\,{\rm GeV}(\mu/100~{\rm
  GeV})^{2/3}$ and $8\times 10^{7}\,{\rm GeV}(m_{\tilde{g}}/1~{\rm
  TeV})^{2/3}$~\cite{Ibanez:1992aj,Fong:2020fwk}. As temperature gets
down to $T\sim M_{1,2}$, the lepton number is transported by the
interactions that are in equilibrium. In the MSSM, there are 18
independent fields and 13 types of interactions~\cite{Fong:2010qh}.
Five U(1) charges are anomaly-free, which are hypercharge,
$\Delta_\alpha \equiv B/3-L_\alpha$ ($\alpha=e,\mu,\tau$), and ${\cal
  R}$ defined in \cite{Fong:2010qh}. Here $B$ and $L_\alpha$ are the
baryon number and the lepton number of each flavor. It is worth noting
that ${\cal R}$ is different from the $R$-symmetry of the
supersymmetric model. The rest of 13 U(1) charges are broken when the
all interactions are in equilibrium. We can take a convenient linear
combination of the charges, which are broken one by one when an
interaction enters in equilibrium as the temperature gets
lower.\footnote{We take into account the neutrino Yukawa interactions
later.  Since the right-handed neutrinos are gauge singlets, their
scalar and fermionic components are independent. Therefore, two
additional degrees of freedom with two interaction terms, i.e.,
Majorana masses and the neutrino Yukawas, lead to no additional U(1)
and the five anomaly free U(1)s are kept unbroken.  Our model has two
new fields $S_\pm$ with one Yukawa interaction. Therefore a U(1)
should appears, which corresponds to the gauged U(1) in our setup.
After inflation, $S_+$ gets the VEV and the gauged U(1) is broken. At
the same time $S_+$ obtains a mass of $\sqrt{\xi}\sim
\order{10^{15}}\,{\rm GeV}$, which is integrated out in the energy
scale we are interested in. The mass scale of $S_-$ is the order of
the inflaton mass. Therefore, it is also integrated out below the
energy scale of the reheating, as well as the inflaton field.  } The
U(1) charges are listed in Appendix~\ref{app:charges}. The crucial
point is that at the decay of the inflaton non-zero conserved charges
are created in addition to $\Delta_\alpha$. As a consequence, the
lepton number stored in the conserved charges escapes from the
wash-out by $\hat{N}_1$ and $\hat{N}_2$.\footnote{ The chemical
potential of the right-handed neutrinos are Boltzmann-suppressed and
irrelevant for the conditions of the
equilibrium~\cite{Engelhard:2006yg,Bertuzzo:2010et,Fong:2010qh}, for
instance, given as
$\mu_{\ell_\alpha}+\mu_{\tilde{H}_u}+\mu_{\tilde{g}}=0$. See also
Appendix~\ref{app:B-L}. } On the other hand, $\Delta_\alpha$ are
affected by the wash-out effect.  Therefore, we have two contributions
to the $B-L$, the conserved charges and the relic that survives the
wash-out.

In the present case, left-handed (s)leptons and up-type Higgsinos
(Higgses) are produced by the inflaton decay. Additionally, ${\cal R}$
and $R^\chi_{3q}$ are expected to be generated. This is due to the $R$
symmetry breaking during the coherent inflaton oscillation. Therefore
the inflaton decay gives an initial condition of the chemical
potentials of the conserved charges are written as
\begin{align}
 \mu_{Q^i} =(\mu_R,-r_e\mu_{L^{\rm dec}},-r_\mu\mu_{L^{\rm
     dec}},-r_\tau\mu_{L^{\rm dec}},\mu_R,\mu_R,\mu_R,\mu_R,\mu_R)\,,
\end{align}
where $Q^i=({\cal R},\Delta_e,\Delta_\mu,\Delta_\tau,
R^\chi_{3u},R^\chi_{3d},R^\chi_{3s},R^\chi_{3c},R^\chi_{3b})$ and the
other chemical potentials are zero.  $\mu_{L^{\rm dec}}$ is the
chemical potential of the total lepton number produced at the inflaton
decay, i.e., $n_L|_{\rm dec} = \mu_{L^{\rm dec}} T^2/6$ and $r_\alpha$
are the fractions of each flavor. $\mu_R$ is the chemical potential
of $R$ number.

A finite value of $\mu_R$ may come from inflation or the coherent
oscillation of the inflaton field. In the VEV $\expval{\phi}$ or the
variance $\sqrt{\expval{\phi^2}}$ of the inflaton field induces
additional one-loop diagram with $R$-breaking intermediate states
appears, which leads to an asymmetry of $R$. With the variance, for
instance, the asymmetry is estimated to be suppressed by at least
$(\hat{y}\hat{y}^\dagger)_{33} \expval{\phi^2}/m_\phi^2$, compared
with $\epsilon_\phi$. Here we found the same suppression factor
$\Lambda/m_\phi$ from the imaginary part of the Yukawa couplings as
$\epsilon_\phi$. Then, it is $\order{10^{-10}}$ suppression in our
target parameter space.\footnote{We use an estimation of the variance
$\sqrt{\expval{\phi^2}}\sim (\hat{y}\hat{y}^\dagger)_{33}/(4\pi)M_{\rm
  pl}\sim 10^{11}\,{\rm GeV}\times
\left[(\hat{y}\hat{y}^\dagger)_{33}/10^{-6} \right]$, where $M_{\rm
  pl}$ is the reduced Planck mass. } Therefore it can be ignored in
our current study and we omit the contribution from $\mu_R$ in the
discussion below.  On the other hand, it may be an importnat
contribution to the baryon assymmetry if
$(\hat{y}\hat{y}^\dagger)_{33}\sim 1$. In that case, we need to take
into account the non-perturbative decay of inflaton during the
coherent oscillation.  Or if we consider a different type of the
seesaw mechanism, the suppression factor $\Lambda/m_\phi$ may be
irrelevant to boost the asymetric parameter.  We will leave these
possibilities for the future research.

Let us consider $M_{1,2}\sim 10^{9}$--$10^{10}\,{\rm GeV}$. Using
Eq.\,\eqref{eq:B-L_T1} in Appendix~\ref{app:B-L}, we get
\begin{align}
  \mu_{Q^{B-L}}=
  \frac{120251}{148420}\mu_{Q^{\Delta_0}}
  \simeq
  0.81\mu_{Q^{\Delta_0}}\,,
    \label{eq:mu_B-L_1}
\end{align}
Referring to Ref.\,\cite{Bertuzzo:2010et}, $\mu_{Q^{\Delta_0}}$ can be
obtained as follows. At the temperature the QCD and electroweak
sphaleron processes, $b$, $\tau$, and $c$ Yukawa interactions are in
equilibrium. Due to the $\tau$ Yukawa interaction, $\ket*{\ell_\tau}$
component in $\ket*{\ell_\phi}$ is washed out since $K_{I}^\tau \gg 1$
for $I=1,2$, where $K_{I}^\alpha \equiv
K_I|\braket*{\ell_\alpha}{\ell_I}|^2$
($\alpha=e,\mu,\tau$).\footnote{We consider no fine-tuning in
$K_I^\alpha$. Namely, we consider $K_I^\alpha\gg 1$. } Let us call
$\ket*{\ell_\tau^\perp}$ as a state orthogonal to
$\ket*{\ell_{\tau}}$.  Next, we decompose $\ket*{\ell_{\tau}^\perp}$
by states $\ket*{\ell_{I}^{\tau_\perp}}$ and
$\ket*{\ell_{I_\perp}^{\tau_\perp}}$ ($I=1,2$); the former is
$\ket*{\ell_I}$ projected on to a plane perpendicular to
$\ket*{\ell_\tau}$ and the latter is one which is orthogonal to
$\ket*{\ell_{I}^{\tau_\perp}}$ in that plane. Then
$\ket*{\ell_{I}^{\tau_\perp}}$ component is washed out due to
$K_{I}^{e}+K_{I}^{\mu} \gg 1$. To summarize, using the decomposition
\begin{align}
  \ket*{\ell_{\phi}}&=
  C_{\phi\tau}\ket*{\ell_\tau}+C_{\phi\tau^\perp}\ket*{\ell_\tau^\perp}\,,
  \label{eq:l_phi}
  \\
  \ket*{\ell_\tau^\perp}&=C_{\tau^\perp 2}\ket*{\ell_{2}^{\tau_\perp}}
  +C_{\tau^\perp 2^\perp}\ket*{\ell_{2_\perp}^{\tau_\perp}}\,,
  \label{eq:l_tau^perp}
  \\
  \ket*{\ell_{2_\perp}^{\tau_\perp}}&=C_{2^\perp 1}\ket*{\ell_{1}^{\tau_\perp}}
  +C_{2^\perp 1^\perp}\ket*{\ell_{1_\perp}^{\tau_\perp}}\,,
  \label{eq:l2_perp^perp}
\end{align}
a fraction $|C_{\phi\tau^\perp}C_{\tau^\perp 2^\perp}C_{2^\perp
  1^\perp}|^2$ of the $B-L$ produced by the inflaton decay survives,
i.e., $\mu_{Q^{\Delta_0}}=|C_{\phi\tau^\perp}C_{\tau^\perp
  2^\perp}C_{2^\perp 1^\perp}|^2 \mu_{L^{\rm
    dec}}$~\cite{Bertuzzo:2010et}. $\mu_{Q^{\Delta_0}}$ depends on the
details of the model parameters, such as $y_{ij}$, $M_{ij}$, and
$\lambda_i$.

If we consider $M_{1,2}\sim 10^{6}$--$10^{9}\,{\rm GeV}$, then all
charged leptons are distinguished. Then all $B-L_\alpha$ components
are washed out and no sufficient $B-L$ is obtained to explain the
observed baryon number.

\begin{figure}[t]
  \begin{center}
    \includegraphics[scale=0.3]{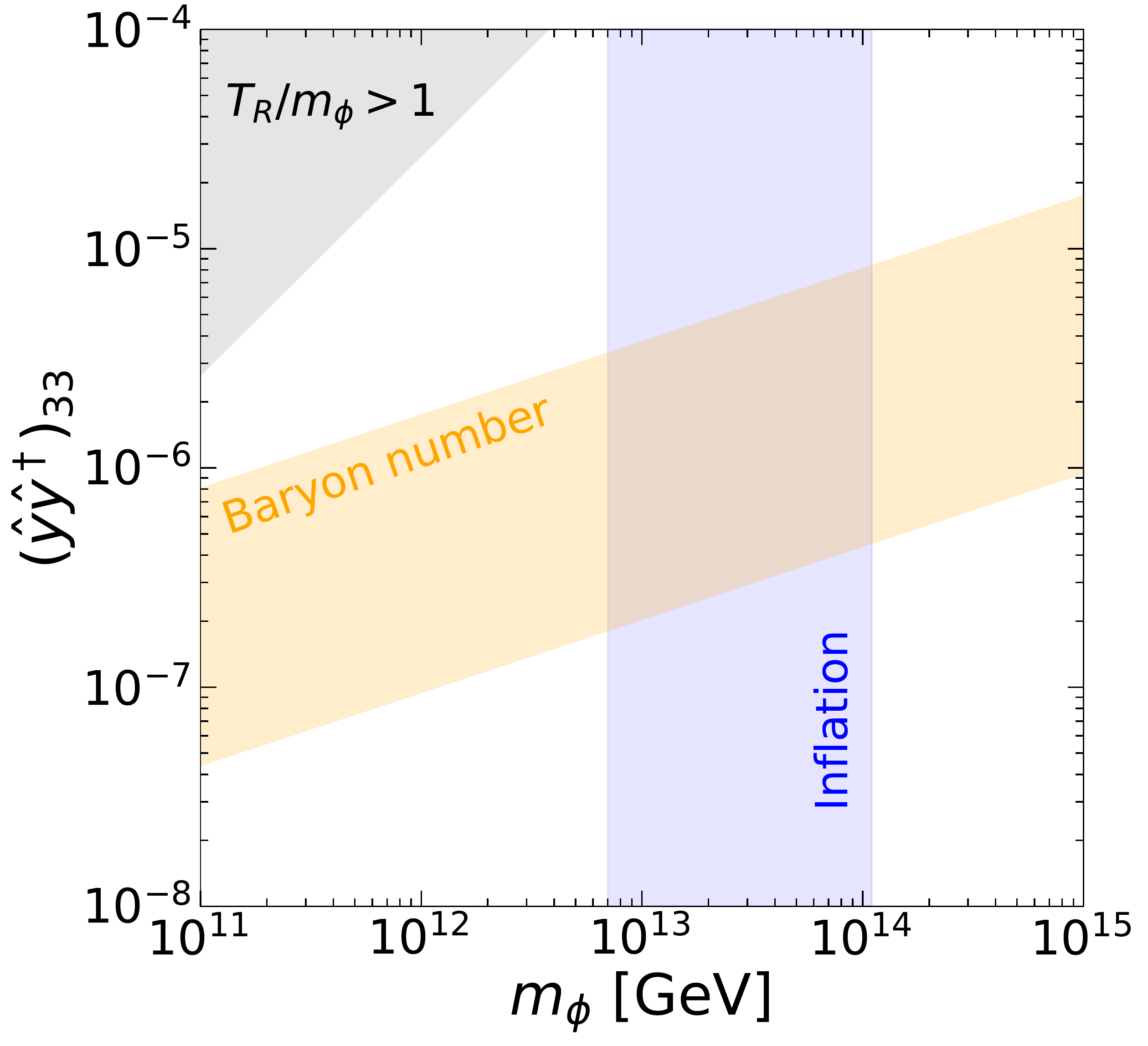}
  \end{center}
  \caption{Parameter region which is consistent with the observed
    baryon asymmetry (orange, ``Baryon number''). Here we take
    $ad=0.01$ and $0.8$ and use $1\sigma$ region of the observed
    baryon number given in Eq.\,\eqref{eq:eta_obs}. Shaded region on
    the top-left corner indicate $T_R/m_{\phi}>1$ (gray). We also plot
    the inflaton mass region \eqref{eq:inflaton_mass_range}, which is
    preferred the superconformal subcritical hybrid
    inflation~\cite{Ishiwata:2018dxg,Gunji:2021zit} (blue,
    ``Inflation'').}
  \label{fig:eta}
\end{figure}

Finally at the stage {\it d)}, the $B-L$ is converted to the baryon
number as $Y_{B}=(10/31)Y_{B-L}$. Combining Eqs.\,\eqref{eq:YLdec},
\eqref{eq:TR}, \eqref{eq:epsilon_phi}, and \eqref{eq:mu_B-L_1}, we get
\begin{align}
  Y_B \simeq 1.2\times 10^{-11}
  \left(\frac{ad}{0.01}\right)
  \left(\frac{m_\phi}{10^{13}\,{\rm GeV}}\right)^{-1/2}
  \left(\frac{(\hat{y}\hat{y}^\dagger)_{33}}{10^{-6}}\right)^{3/2}\,.
\end{align}
Here we take into account the spectator effects and the wash-out
effect by introducing a coefficient $d$, which range from about $0.04$
to $0.8$. The result is plotted on
($(\hat{y}\hat{y}^\dagger)_{33},m_\phi$) plane in
Fig.\,\ref{fig:eta}. Here we also indicate the region where inflation
induced by $\tilde{N}'_3$ predicts the spectral index and the
tensor-to-scalar ratio that are consistent with the Planck observation
based on Refs.\,\cite{Ishiwata:2018dxg,Gunji:2021zit} (see
Appendix~\ref{app:m_phi} for details):
\begin{align}
  0.7\times 10^{13}\,{\rm GeV}<m_\phi<11\times 10^{13}\,{\rm GeV}\,.
  \label{eq:inflaton_mass_range}
\end{align}
Therefore, the observed baryon number is obtained after the successful
inflation in the region $(\hat{y}\hat{y}^\dagger)_{33}\sim
10^{-7}$--$10^{-6}$ and $m_\phi\sim 10^{13}\,{\rm GeV}$. This value of
the neutrino Yukawa coupling is desirable for the reason discussed
below. Let us say that all $y_{ij}$ have the same order.  Then we
obtain $(\hat{y}\hat{y}^\dagger)\sim m_i \Lambda/\expval{H_u}^2\sim
10^{-6}(\Lambda/10^{10}\,{\rm GeV})$ from the observed neutrino
masses.  This is consistent with the assumption \eqref{eq:M_{1,2}}.
In addition, the baryon number is predicted to behave as $Y_B\propto
\Lambda^{3/2}$, which means that $\Lambda\sim 10^{10}\,{\rm GeV}$,
i.e. $M_{1,2}\sim 10^{10}\,{\rm GeV}$, is required to get the observed
number. Therefore, in a case where all $y_{ij}$ are the same order,
the observed baryon asymetry can be obtained in the setup of this
hybrid inflation model that is preferred by both the Planck
observation and the neutrino masses.

\section{Conclusion}
\label{sec:conclusion}
\setcounter{equation}{0}

We consider a model of supersymmetric hybrid inflation and study the
reheating and baryogenesis after inflation. The model consists of
three right-handed neutrinos $N_i$ with the Majorana masses and two
fields $S_{\pm}$ that are charged under a gauged U(1). A scalar
component of a linear combination of the $N_i$ plays the role of the
inflaton, while the $S_+$ is the waterfall field. We focus on a case
where the inflation lasts below the critical point value, called
subcritical hybrid inflation. The inflaton mass should be $[0.7,\,
  11]\times 10^{13}\,{\rm GeV}$ from the observations of the scalar
spectral index and tensor-to-scalar ratio by the Planck
collaboration. In addition, the scale $\Lambda$ of the Majorana masses
needs to be smaller than $\order{10^{10}}\,{\rm GeV}$ in order not to
disturb the inflationary dynamics. Therefore there is a hierarchy
between the inflaton mass and the Majorana mass scale. In addition,
the VEV of the waterfall field results in Dirac mass terms for the
$N_i$ and $S_-$.

After inflation, the inflaton decays to reheat the universe and at the
same time several conserved quantities, including $B/3-L_\alpha$, are
provided. The conserved charges are broken and transported under the
equilibrium conditions as the temperature drops down. Furthermore
$B/3-L_\alpha$ suffers from the wash-out due to the lighter
right-handed (s)neutrinos. The wash-out is inevitably strong because
of the observed neutrino masses and the special structure, i.e., Dirac
and Majorana type, of the mass matrix of the right-handed neutrinos.
In spite of the strong wash-out, a part of $B/3-L_\alpha$ can survive
for $\Lambda\sim 10^{9}$--$10^{10}$\,GeV.  Below that scale, though
there are conserved quantities, such as ${\cal R}$, they have
negligible contributions to the baryon number.  Consequently, we found
a successful baryogenesis for $10^{9}\,{\rm GeV}\lesssim
\Lambda\lesssim 10^{10}\,{\rm GeV}$.

\section*{Acknowledgments}
We thank Chee Sheng Fong for fruitful discussions. This work is
supported by JST SPRING, Grant No. JPMJSP2135 (YG), JSPS KAKENHI Grant
No. JP18H05542, JP20H01894, and JSPS Core-to-Core Program Grant
No. JPJSCCA20200002 (KI).

\appendix

\section{The interactions and charges}
\setcounter{equation}{0} 
\label{app:charges}

We construct a set of U(1) charges in the MSSM based on the technique
given in Ref.\,\cite{Domcke:2020quw}. In our study we ignore the $\mu$
term for Higgses and the masses of gauginos by assuming $\mu$ and the
supersymmetry breaking scale smaller than $\order{10^{9}}$\,{\rm GeV}.

In the MSSM, we have the following relevant
fields~\cite{Ibanez:1992aj,Fong:2010qh,Fong:2012buy}:
\begin{align}
  f&=(e,\mu,\tau,\ell_e,\ell_\mu,\ell_\tau,
  u,c,t,d,s,b,Q_1,Q_2,Q_3,\tilde{H}_u,\tilde{H}_d,\tilde{g})\,,
  \\
  b&=(\tilde{e},\tilde{\mu},\tilde{\tau},
  \tilde{\ell}_e,\tilde{\ell}_\mu,\tilde{\ell}_\tau,
  \tilde{u},\tilde{c},\tilde{t},\tilde{d},\tilde{s},\tilde{b},
  \tilde{Q}_1,\tilde{Q}_2,\tilde{Q}_3, H_u,H_d,g)\,,
\end{align}
where $f$ are fermions and $b$ indicates their bosonic partners.
$e,\mu,\tau$, $u,c,t$, $d,s,b$ are right-handed fields, and the rest
are left-handed fields. The gauge interactions are equilibrium and all
gauginos have the same chemical potential, denoted as $\tilde{g}$. 
The number density asymmetries are given by their chemical potentials
$\mu_i$ ($i=f,b$) as
\begin{align}
  n_i-\bar{n}_i=g_i\mu_i T^2/6\,,
\end{align}
where $g_i$ are the multiplicities defined by
\begin{align}
  g_f &= (1, 1, 1, 2, 2, 2, 3, 3, 3, 3, 3, 3, 6, 6, 6, 2, 2, 12)\,,
  \label{eq:gf}
  \\
  g_b &= (2, 2, 2, 4, 4, 4, 6, 6, 6, 6, 6, 6, 12, 12, 12, 4, 4, 0)\,.
  \label{eq:gb}
\end{align}
Due to the gauge interactions, the chemical potentials of bosonic
partners are given by $\mu_b=\mu_f+\mu_{\tilde{g}}$ and
$\mu_f-\mu_{\tilde{g}}$ for left-handed and right-handed fields,
respectively. Following
Refs.\,\cite{Fong:2010qh,Domcke:2020quw},\footnote{See also
Ref.\,\cite{Fong:2015vna}.} we introduce the interaction vectors:
\begin{align}
  y^t&=(0, 0, 0, 0, 0, 0, 0, 0, -1, 0, 0, 0, 0, 0, 1, 1, 0, 1)\,,
  \label{eq:yt}
  \\
  y^{\rm SS}&=(0, 0, 0, 0, 0, 0, -1, -1, -1, -1, -1, -1, 2, 2, 2, 0, 0, 6)\,,
  \\
  y^{\rm WS}&=(0, 0, 0, 1, 1, 1, 0, 0, 0, 0, 0, 0, 3, 3, 3, 1, 1, 4)\,,
  \\
  y^{b}&=(0, 0, 0, 0, 0, 0, 0, 0, 0, 0, 0, -1, 0, 0, 1, 0, 1, 1)\,,
  \\
  y^\tau&=(0, 0, -1, 0, 0, 1, 0, 0, 0, 0, 0, 0, 0, 0, 0, 0, 1, 1)\,,
  \\
  y^{Q_{23}}&=(0, 0, 0, 0, 0, 0, 0, 0, 0, 0, 0, 0, 0, -1, 1, 0, 0, 0)\,,
  \\
  y^c&=(0, 0, 0, 0, 0, 0, 0, -1, 0, 0, 0, 0, 0, 1, 0, 1, 0, 1)\,,
  \\
  y^\mu&=(0, -1, 0, 0, 1, 0, 0, 0, 0, 0, 0, 0, 0, 0, 0, 0, 1, 1)\,,
  \\
  y^s&=(0, 0, 0, 0, 0, 0, 0, 0, 0, 0, -1, 0, 0, 1, 0, 0, 1, 1)\,,
  \\
  y^{Q_{12}}&=(0, 0, 0, 0, 0, 0, 0, 0, 0, 0, 0, 0, -1, 1, 0, 0, 0, 0)\,,
  \\
  y^d&=(0, 0, 0, 0, 0, 0, 0, 0, 0, -1, 0, 0, 1, 0, 0, 0, 1, 1)\,,
  \\
  y^u&=(0, 0, 0, 0, 0, 0, -1, 0, 0, 0, 0, 0, 1, 0, 0, 1, 0, 1)\,,
  \\
  y^e&=(-1, 0, 0, 1, 0, 0, 0, 0, 0, 0, 0, 0, 0, 0, 0, 0, 1, 1)\,.
  \label{eq:ye}
\end{align}
`SS' and `WS' are strong and weak sphaleron processes, respectively,
and the others are from the Yukawa interactions.\footnote{$y^{Q_{23}}$
and $y^{Q_{12}}$ can be replaced by $y^{sb}=(0, 0, 0, 0, 0, 0, 0, 0,
0, 0, -1, 0, 0, 0, 1, 0, 1, 1)$ and $y^{ds}=(0, 0, 0, 0, 0, 0, 0, 0,
0, -1, 0, 0, 0, 1, 0, 0, 1, 1)$, respectively. } Using the interaction
vectors, the equilibrium condition is given by $\mu_f\vdot y^{int}=0$
(${int}=t,{\rm SS},{\rm WS},\cdots$). For example, $\mu_f\vdot
y^{t}=-\mu_t+\mu_{Q_3}+\mu_{\tilde{H}_u}+\mu_{\tilde{g}}=0$ etc. To
describe the transportation of the chemical potentials, we introduce a
set of charges for fermions $f$ based on
Refs.\,\cite{Domcke:2020quw,Fong:2010qh}. There are five charges which
are anomaly free and conserved under the interactions listed above:
\begin{align}
  q^{Y}_f &=
  \mbox{\footnotesize $
    (-1, -1, -1, -1/2, -1/2, -1/2, 2/3, 2/3, 2/3, -1/3, -1/3, -1/3, 1/6, 
1/6, 1/6, 1/2, -1/2, 0)$}\,,
  \\
  q^{\Delta_e}_f  &=(-1, 0, 0, -1, 0, 0,
  1/9, 1/9, 1/9, 1/9, 1/9, 1/9, 1/9, 1/9, 1/9, 0, 0, 0)\,,
  \\
  q^{\Delta_\mu}_f  &=(0, -1, 0, 0, -1, 0,
  1/9, 1/9, 1/9, 1/9, 1/9, 1/9, 1/9, 1/9, 1/9, 0, 0, 0)\,,
  \\
  q^{\Delta_\tau}_f  &=(0, 0, -1, 0, 0, -1,
  1/9, 1/9, 1/9, 1/9, 1/9, 1/9, 1/9, 1/9, 1/9, 0, 0, 0)\,,
  \\
  q^{\cal R}_f  &=
  \mbox{\footnotesize $
  (2, 2, 2, 0, 0, 0, -4/9, -4/9, -4/9, 14/9, 14/9, 
    14/9, -4/9, -4/9, -4/9, -1, 1, 1)
    $}\,,
\end{align}
corresponding to hypercharge, $\Delta_\alpha=B/3-L_\alpha$
($\alpha=e,\mu,\tau$) and ${\cal R}$ introduced in
Ref.\,\cite{Fong:2010qh}.\footnote{This ${\cal R}$ is different from
the $R$-symmetry. As a reference, $R$-charges of the fermions are
given as
$q^R_f=(-1,-1,-1,-1,-1,-1,3,3,3,3,3,3,-1,-1,-1,3,-1,1)$~\cite{Fong:2010qh}.}
The rest charges are broken at the onset of each interactions from
\eqref{eq:yt} to \eqref{eq:ye}:
\begin{align}
  q^t_f &=(0, 0, 0, 0, 0, 0, 0, 0, 1, 0, 0, 0, 0, 0, 0, 0, 0, 0)\,,
  \label{eq:qt}
  \\
  q^{u}_f &=(0, 0, 0, 0, 0, 0, 1, 0, 0, 0, 0, 0, 0, 0, 0, 0, 0, 0)\,,
  \\
  q^B_f &=(0, 0, 0, 0, 0, 0,
  1/3, 1/3, 1/3, 1/3, 1/3, 1/3, 1/3, 1/3, 1/3, 0, 0, 0)\,,
  \\
  q^{R_{3b}^{\chi}}_f &=
  (5, 5, 5, 2, 2, 2, -3, -3, -3, 2, 2, 5, -1, -1, -1, -3, 2, 1)\,,
  \\
  q^\tau_f &=(0, 0, 1, 0, 0, 0, 0, 0, 0, 0, 0, 0, 0, 0, 0, 0, 0, 0)\,,
  \\
  q^{B_1-B_2}_f &
  =(0, 0, 0, 0, 0, 0, 1/3, -1/3, 0, 1/3, -1/3, 0, 1/3, -1/3, 0, 0, 0, 0)\,,
  \\
  q^{R_{3c}^{\chi}}_f &=
  (5, 5, 5, 2, 2, 2, -3, 0, -3, 2, 2, 2, -1, -1, -1, -3, 2, 1)\,,
  \\
  q^\mu_f &=(0, 1, 0, 0, 0, 0, 0, 0, 0, 0, 0, 0, 0, 0, 0, 0, 0, 0)\,,
  \\
  q^{R_{3s}^{\chi}}_f &
  =(5, 5, 5, 2, 2, 2, -3, -3, -3, 2, 5, 2, -1, -1, -1, -3, 2, 1)\,,
  \\
  q^{2B_1-B_2-B_3}_f&=
  \mbox{\footnotesize $
    (0, 0, 0, 0, 0, 0, 2/3, -1/3, -1/3, 2/3, -1/3, -1/3, 2/3,
    -1/3, -1/3, 0, 0, 0)$}\,,
  \\
  q^{R_{3d}^{\chi}}_f&=
  (5, 5, 5, 2, 2, 2, -3, -3, -3, 5, 2, 2, -1, -1, -1, -3, 2, 1)\,,
  \\
  q^{R_{3u}^{\chi}}_f&=
  (5, 5, 5, 2, 2, 2, 0, -3, -3, 2, 2, 2, -1, -1, -1, -3, 2, 1)\,,
  \\
  q^e_f&=(1, 0, 0, 0, 0, 0, 0, 0, 0, 0, 0, 0, 0, 0, 0, 0, 0, 0)\,.
  \label{eq:qe}
\end{align}
The charge assignments for bosons are given by $q^i_b=q^i_f+1$ and
$q^i_f-1$ ($i=Y,\Delta_e,\Delta_\mu,\cdots$) for left-handed and
right-handed fields, respectively, for ${\cal R}$ and $R_{3q}^{\chi}$
($q=b,c,s,d,u$), and they are the same as fermions for the others. We
have introduced a modified version of charge, denoted as
$R_{3q}^{\chi}$, based on the {\it chiral} $R_3$ charge in
Ref.\,\cite{Fong:2010qh}. With above definition and
Eqs\,.\eqref{eq:gf} and \eqref{eq:gb}, the chemical potentials of the
conserved charges are given by
\begin{align}
  \mu_{Q^i} = (q^i_f \circ g_f)\vdot \mu_f +(q^i_b\circ g_b)\vdot \mu_b\,,
\end{align}
where $\circ$ denotes the entrywise Hadamard product. We always impose
$\mu_{Q^Y}=0$. With the chemical potentials $\mu_{Q^i}$, the asymmetry number
density for the conserved charges are given by
\begin{align}
  n^{Q^i}-\bar{n}^{Q^i}=\mu_{Q^i}T^2/6\,.
\end{align}
For example, the asymmetry of $B-L$ is given by $\sum_\alpha
\mu_{Q^{\Delta_\alpha}} T^2/6$.

\section{$B-L$ at the wash-out regime}
\setcounter{equation}{0} 
\label{app:B-L}

With the interactions and the conserved charges introduced in the
previous section, we can calculate the $B-L$ for a given set of the
equilibrium conditions. Here we give several temperature regime from
$10^{10}\,{\rm GeV}$ to $10^{5}\,{\rm GeV}$. Though we focus on the
range $[10^{7}\,{\rm GeV},10^{10}\,{\rm GeV}]$, we give the result as
a reference. In out study, we consider $\tan \beta $, the ratio of the
VEV of the up-type and down-type Higgs, is $ \sim 1$ and adopt the
equilibrium temperatures of the relevant interactions given in
Refs.\,\cite{Domcke:2020kcp,Fong:2020fwk}. To check our calculation,
we compute $A^\ell$, $C^{\tilde{g}}$, $C^{\tilde{H}_u}$, and
$C^{\tilde{H}_d}$ in Ref.\,\cite{Fong:2010qh} and obtain consistent
results from Eqs.~(2.40)-(2.45) in the literature, except for
Eq.~(2.41).\footnote{We thank Chee Sheng Fong for confirming this
point. }

\noindent \underline{{\bf (i)} $T\sim 10^{9}$--$10^{10}\,{\rm GeV}$}:
$t$, $b$, $c$, $\tau$ Yukawa interactions and strong, weak sphaleron
processes are in equilibrium. In this case, $\tau$ (including
$\ell_\tau$) is distinguished. On the other hand, a linear combination
of $\ell_e$ and $\ell_\mu$, which are tentatively denoted as $\ell_e'$
and $\ell_\mu'$, are disentangled if the interaction with $\hat{N}_1$ and
$\hat{N}_2$ are in equilibrium, which will be discussed later.

It is straightforward to compute the chemical potentials of
$\ell_{e}'$, $\ell_{\mu}'$, $\ell_\tau$, $\tilde{H}_u$, and
$\tilde{g}$ in terms of $\mu_{Q^i}$
($i=\Delta_{e'},\Delta_{\mu'},\Delta_\tau,{\cal
  R},e',\mu',R^\chi_{3u},R^\chi_{3d},2B_1-B_2-B_3,R^\chi_{3s},B_1-B_2$).
The coefficients of $\mu_{Q^i}$ are given by
\begin{align}
  \mu_{\ell_{e}'}:~&
  \mbox{ \small
    $-\frac{432337}{3164482},\frac{142615}{4746723},
    \frac{100513}{4746723},-\frac{132149}{9493446},
    -\frac{1329629}{9493446},\frac{42102}{1582241},
    \frac{4943}{1582241},
    \frac{297}{93073},-\frac{10045}{1582241},\frac{10204}{4746723},
    \frac{7653}{1582241}$}
  \\
    \mu_{\ell_{\mu}'}:~&
    \mbox{ \small
      $\frac{142615}{4746723},-\frac{432337}{3164482},
      \frac{100513}{4746723},-\frac{132149}{9493446},
      \frac{42102}{1582241},-\frac{1329629}{9493446},
      \frac{4943}{1582241},\frac{297}{93073},
      -\frac{10045}{1582241},\frac{10204}{4746723},
      \frac{7653}{1582241}
      $
    }
  \\
  \mu_{\ell_\tau}:~&
  \mbox{\small
    $\frac{167309}{9493446},\frac{167309}{9493446},
    -\frac{166914}{1582241},-\frac{273605}{9493446},
    \frac{341897}{9493446},\frac{341897}{9493446},
    \frac{11213}{3164482},
    \frac{6013}{558438},-\frac{56717}{3164482},
    \frac{15168}{1582241},\frac{34128}{1582241}
    $
  }
  \\
  \mu_{\tilde{H}_u}:~&
  \mbox{\small
    $
    -\frac{94921}{3164482},-\frac{94921}{3164482},
    -\frac{52238}{1582241},\frac{46760}{1582241},
    \frac{28665}{3164482},\frac{28665}{3164482},
    -\frac{203327}{9493446},-\frac{13679}{1675314},
    \frac{72645}{3164482},-\frac{14608}{14240169},
   -\frac{3652}{1582241}
    $
  }
  \\
  \mu_{\tilde{g}}:~&
  \mbox{\small
    $-\frac{17141}{1582241},-\frac{17141}{1582241},
    -\frac{17983}{1582241},\frac{45461}{3164482},
    \frac{2526}{1582241},\frac{2526}{1582241},
    \frac{1424}{1582241},\frac{332}{279219},
    -\frac{3534}{1582241},\frac{4220}{4746723},
    \frac{3165}{1582241}
    $
        }\,.
\end{align}
Now we take into account the strong wash-out effect. The chemical
potentials after the wash-out are given by $\mu_{\ell_1} +
\mu_{\tilde{H}_u}+\mu_{\tilde{g}}=0$ and $\mu_{\ell_2} +
\mu_{\tilde{H}_u}+\mu_{\tilde{g}}=0$.  Here it should be noted that a
charged lepton $\ell^{\tau_\perp}_{1_\perp}$ in
Eq.\,\eqref{eq:l2_perp^perp} does not couple to $\hat{N}_1$ and $\hat{N}_2$ and
the corresponding charge is conserved and the others are
broken. However, it is not trivial to extract such a state from the
equilibrium equations and solve them analytically. Following
Ref.\,\cite{Engelhard:2006yg}, we approximately estimate the wash-out
effect by solving $\mu_{\ell_\tau} +
\mu_{\tilde{H}_u}+\mu_{\tilde{g}}=0$ and $\mu_{\ell_\mu'} +
\mu_{\tilde{H}_u}+\mu_{\tilde{g}}=0$ and identify $\ell_e'$ as
$\ell^{\tau_\perp}_{1_\perp}$. A crucial point is that the neutrino
Yukawa interactions do not break the other symmetries. Namely,
$\Delta_0,{\cal R},e',\mu',R^\chi_{3u},R^\chi_{3d},2B_1-B_2-B_3,
R^\chi_{3s}$, and $B_1-B_2$ are unbroken. Here we rewrite
$\Delta_{e'}$ as $\Delta_0$.  Therefore, in the basis $\mu_{Q^i}$
($i=\Delta_0,{\cal R},e',\mu',R^\chi_{3u},R^\chi_{3d},2B_1-B_2-B_3,
R^\chi_{3s},B_1-B_2$) the coefficients to give chemical
potential for $B-L$ are given by
\begin{align}
  \mu_{Q^{B-L}}:~\mbox{\small
    $\frac{120251}{148420},
    \frac{3491}{14842},\frac{67889}{148420},
    -\frac{26181}{74210},-\frac{27463}{148420},
    \frac{1801}{445260},\frac{12831}{148420},
    \frac{7316}{111315},\frac{5487}{37105}
   $
   }\,.
  \label{eq:B-L_T1}
\end{align}

\noindent \underline{{\bf (ii)} $T\sim 10^{6}$--$10^{9}\,{\rm GeV}$}:
In addition to the previous case, $\mu$, $s$, $Q_{23}$ Yukawa
interactions are in equilibrium.  Then all charged leptons are
disentangled. The results are
($i=\Delta_{e},\Delta_{\mu},\Delta_\tau,{\cal
  R},e,R^\chi_{3u},R^\chi_{3d},2B_1-B_2-B_3$).  The coefficients of
$\mu_{Q^i}$ are given by
\begin{align}
  \mu_{\ell_{e}}:~&
  \mbox{ \small
    $
    -\frac{49715}{369014},\frac{13217}{553521},
    \frac{13217}{553521},-\frac{11719}{1107042},
    -\frac{157723}{1107042},\frac{1759}{553521},
    \frac{1015}{553521},-\frac{1387}{369014}
    $
  }
  \\
    \mu_{\ell_{\mu}}:~&
    \mbox{ \small
      $
      \frac{45005}{2214084},-\frac{165299}{1660563},
      \frac{19208}{1660563},-\frac{19111}{1107042},
      \frac{58183}{2214084},\frac{9695}{2214084},
      \frac{5571}{738028},-\frac{3301}{369014}
      $
    }
  \\
  \mu_{\ell_\tau}:~&
  \mbox{\small
    $\frac{45005}{2214084},\frac{19208}{1660563},
    -\frac{165299}{1660563},-\frac{19111}{1107042},
    \frac{58183}{2214084},
    \frac{9695}{2214084},\frac{5571}{738028},
    -\frac{3301}{369014}
    $
  }
  \\
  \mu_{\tilde{H}_u}:~&
  \mbox{\small
    $
    -\frac{21687}{738028},-\frac{6033}{184507},
    -\frac{6033}{184507},\frac{5320}{184507},
    \frac{7335}{738028},-\frac{143473}{6642252},
    -\frac{54887}{6642252},\frac{8265}{369014}
    $
  }
  \\
  \mu_{\tilde{g}}:~&
  \mbox{\small
    $
     -\frac{1975}{184507},-\frac{2019}{184507},
    -\frac{2019}{184507},\frac{5671}{369014},
    \frac{132}{184507},\frac{548}{553521},
    \frac{526}{553521},-\frac{537}{369014}
    $
        }\,.
\end{align}
As the previous case, we solve $\mu_{\ell_\alpha} +
\mu_{\tilde{H}_u}+\mu_{\tilde{g}}=0$ ($\alpha=e,\mu,\tau$) to take
into account the wash-out effect.  In this case all $\Delta_\alpha$ are
broken due to the neutrino Yukawa interactions. Consequently, the
chemical potential for $B-L$ is given by the chemical potentials of
the unbroken charges $\mu_{Q^i}$ ($i={\cal
  R},e,R^\chi_{3u},R^\chi_{3d},2B_1-B_2-B_3$),
\begin{align}
  \mu_{Q^{B-L}}:~\mbox{\small
    $\frac{85794}{198313},-\frac{42771}{198313},
    -\frac{49149}{198313},-\frac{4455}{198313},
    \frac{40203}{198313}
   $
  }\,.
  \label{eq:B-L_T2}
\end{align}

\noindent \underline{{\bf (iii)} $T\sim 10^{5}$--$10^{6}\,{\rm GeV}$}:
Under this temperature only $e$ Yukawa interaction is out of
equilibrium.  The coefficients of $\mu_{Q^i}$
($i=\Delta_{e},\Delta_{\mu},\Delta_\tau,{\cal R},e$) are given by
\begin{align}
  \mu_{\ell_{e}}:~&
  \mbox{ \small
    $
    -\frac{931}{6786},\frac{211}{9657},
    \frac{211}{9657},-\frac{1}{174},-\frac{415}{2886}
    $
  }
  \\
    \mu_{\ell_{\mu}}:~&
    \mbox{ \small
      $
      \frac{113}{6786},-\frac{326}{3219},
      \frac{95}{9657},-\frac{1}{174},\frac{59}{2886}
      $
    }
  \\
  \mu_{\ell_\tau}:~&
  \mbox{\small
    $
    \frac{113}{6786},\frac{95}{9657},
    -\frac{326}{3219},-\frac{1}{174},\frac{59}{2886}
    $
  }
  \\
  \mu_{\tilde{H}_u}:~&
  \mbox{\small
    $
    -\frac{1}{78},-\frac{2}{111},-\frac{2}{111},0,\frac{15}{962}
    $
  }
  \\
  \mu_{\tilde{g}}:~&
  \mbox{\small
    $
    -\frac{1}{87},-\frac{1}{87},-\frac{1}{87},\frac{1}{58}
    $
  }\,.
\end{align}
By solving $\mu_{\ell_\alpha} +
\mu_{\tilde{H}_u}+\mu_{\tilde{g}}=0$ ($\alpha=e,\mu,\tau$), the 
chemical potential
for $B-L$ is given by
\begin{align}
  \mu_{Q^{B-L}}:~\mbox{\small
    $
    \frac{213}{1012},-\frac{261}{1012}
   $
   }\,,
\end{align}
in the basis $\mu_{Q^i}$ ($i={\cal R},e$).

\section{Inflaton mass}
\setcounter{equation}{0} 
\label{app:m_phi}

\begin{figure}[t]
  \begin{center}
    \includegraphics[scale=0.6]{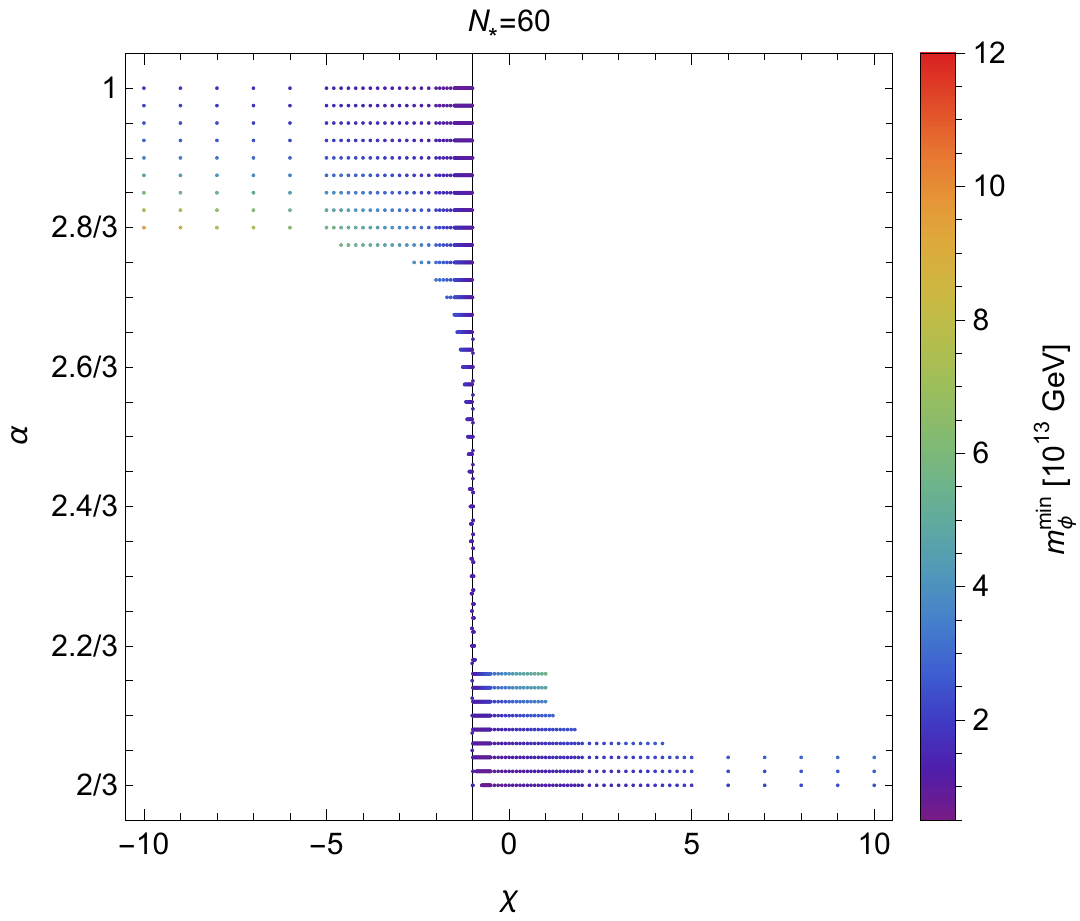}
    \includegraphics[scale=0.6]{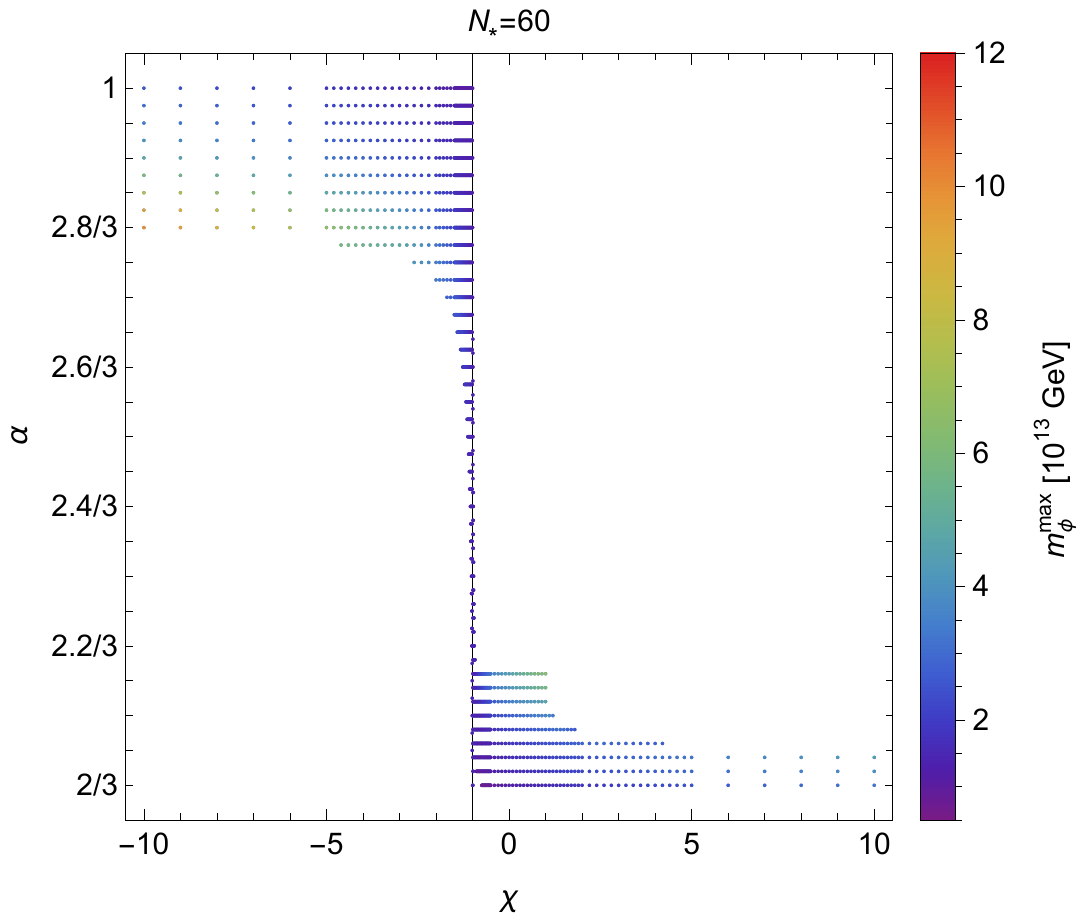}
    \includegraphics[scale=0.6]{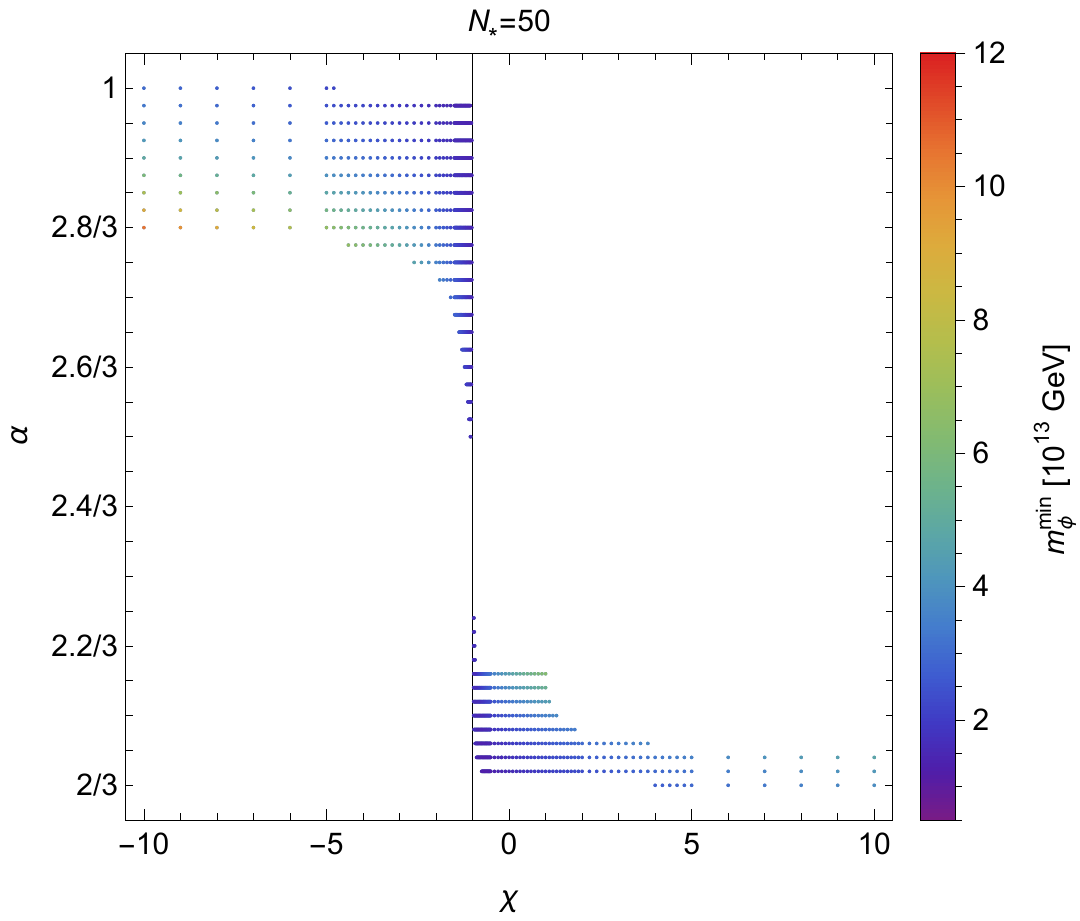}
    \includegraphics[scale=0.6]{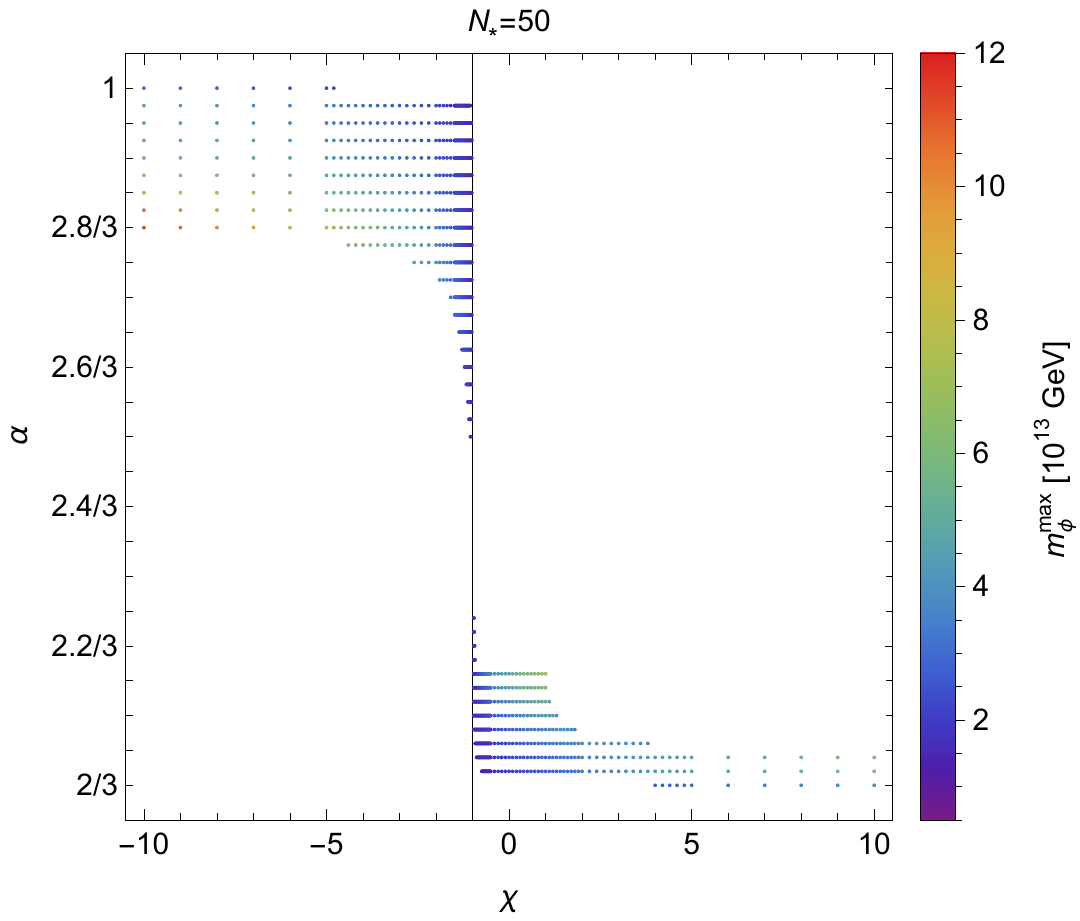}
  \end{center}
  \caption{Minimum (left) and maximum (right) values of the inflaton
    mass for a given set of parameters, $\alpha$ and $\chi$, that
    characterize the symmetry of the K\"ahler potential of the
    model\cite{Gunji:2021zit}. We take the number of $e$-folds as
    $60$ (top) and $50$ (bottom). }
  \label{fig:mphi}
\end{figure}

We use the latest Planck data~\cite{Aghanim:2018eyx,Akrami:2018odb}
\begin{align}
  n_{s}&=0.9649\pm0.0042\,(68\%\,\mathrm{C.L.})\,,
  \\
  r&<0.10\,(95\%\,\mathrm{C.L.})\,,
 \\
  A_s&=2.100\pm0.030\times10^{-9}\,(68\%\,\mathrm{C.L.})\,,
\end{align}
where $n_s$, $r$, and $A_s$ are the spectral index of the scalar mode,
tensor-to-scalar ratio, and the amplitude of the scalar mode.  Based
on the analysis given in Ref.\,\cite{Gunji:2021zit}, we evaluate the
inflaton mass. The result is shown in Fig.\,\ref{fig:mphi}, where the
minimum and maximum values of the inflaton mass is indicated for a
given set of the model parameters. We found the inflaton mass range
that is consistent with the Planck result is
\begin{align}
  0.7\times 10^{13}\,{\rm GeV}<m_\phi<10\times 10^{13}\,{\rm GeV}\,,
\end{align}
for 60 $e$-folds and
\begin{align}
  1.2\times 10^{13}\,{\rm GeV}<m_\phi<11\times 10^{13}\,{\rm GeV}\,,
\end{align}
for 50 $e$-folds.

\bibliography{draft}

\begin{thebibliography}{10}

\bibitem{Aghanim:2018eyx}
Planck, N.~Aghanim {\em et~al.},
\newblock Astron. Astrophys. {\bf 641}, A6 (2020), arXiv:1807.06209.

\bibitem{Fukugita:1986hr}
M.~Fukugita and T.~Yanagida,
\newblock Phys. Lett. B {\bf 174}, 45 (1986).

\bibitem{Minkowski:1977sc}
P.~Minkowski,
\newblock Phys. Lett. B {\bf 67}, 421 (1977).

\bibitem{Yanagida:1979as}
T.~Yanagida,
\newblock Conf. Proc. C {\bf 7902131}, 95 (1979).

\bibitem{Yanagida:1980xy}
T.~Yanagida,
\newblock Prog. Theor. Phys. {\bf 64}, 1103 (1980).

\bibitem{Gell-Mann:1979vob}
M.~Gell-Mann, P.~Ramond, and R.~Slansky,
\newblock Conf. Proc. C {\bf 790927}, 315 (1979), arXiv:1306.4669.

\bibitem{Ramond:1979py}
P.~Ramond,
\newblock {The Family Group in Grand Unified Theories},
\newblock in {\em {International Symposium on Fundamentals of Quantum Theory
  and Quantum Field Theory}}, 1979, arXiv:hep-ph/9809459.

\bibitem{Glashow:1979nm}
S.~L. Glashow,
\newblock NATO Sci. Ser. B {\bf 61}, 687 (1980).

\bibitem{Ishiwata:2018dxg}
K.~Ishiwata,
\newblock Phys. Lett. B {\bf 782}, 367 (2018), arXiv:1803.08274.

\bibitem{Gunji:2021zit}
Y.~Gunji and K.~Ishiwata,
\newblock Phys. Rev. D {\bf 104}, 123545 (2021), arXiv:2104.02248.

\bibitem{Buchmuller:2014rfa}
W.~Buchmuller, V.~Domcke, and K.~Schmitz,
\newblock JCAP {\bf 11}, 006 (2014), arXiv:1406.6300.

\bibitem{Buchmuller:2014dda}
W.~Buchmuller and K.~Ishiwata,
\newblock Phys. Rev. D {\bf 91}, 081302 (2015), arXiv:1412.3764.

\bibitem{Clesse:2010iz}
S.~Clesse,
\newblock Phys. Rev. D {\bf 83}, 063518 (2011), arXiv:1006.4522.

\bibitem{Clesse:2012dw}
S.~Clesse and B.~Garbrecht,
\newblock Phys. Rev. D {\bf 86}, 023525 (2012), arXiv:1204.3540.

\bibitem{Kodama:2011vs}
H.~Kodama, K.~Kohri, and K.~Nakayama,
\newblock Prog. Theor. Phys. {\bf 126}, 331 (2011), arXiv:1102.5612.

\bibitem{Gunji:2019wtk}
Y.~Gunji and K.~Ishiwata,
\newblock JHEP {\bf 09}, 065 (2019), arXiv:1906.04530.

\bibitem{Barbieri:1999ma}
R.~Barbieri, P.~Creminelli, A.~Strumia, and N.~Tetradis,
\newblock Nucl. Phys. B {\bf 575}, 61 (2000), arXiv:hep-ph/9911315.

\bibitem{Abada:2006fw}
A.~Abada, S.~Davidson, F.-X. Josse-Michaux, M.~Losada, and A.~Riotto,
\newblock JCAP {\bf 04}, 004 (2006), arXiv:hep-ph/0601083.

\bibitem{Nardi:2006fx}
E.~Nardi, Y.~Nir, E.~Roulet, and J.~Racker,
\newblock JHEP {\bf 01}, 164 (2006), arXiv:hep-ph/0601084.

\bibitem{Abada:2006ea}
A.~Abada {\em et~al.},
\newblock JHEP {\bf 09}, 010 (2006), arXiv:hep-ph/0605281.

\bibitem{Blanchet:2006be}
S.~Blanchet and P.~Di~Bari,
\newblock JCAP {\bf 03}, 018 (2007), arXiv:hep-ph/0607330.

\bibitem{Antusch:2006cw}
S.~Antusch, S.~F. King, and A.~Riotto,
\newblock JCAP {\bf 11}, 011 (2006), arXiv:hep-ph/0609038.

\bibitem{Buchmuller:2001sr}
W.~Buchmuller and M.~Plumacher,
\newblock Phys. Lett. B {\bf 511}, 74 (2001), arXiv:hep-ph/0104189.

\bibitem{Nardi:2005hs}
E.~Nardi, Y.~Nir, J.~Racker, and E.~Roulet,
\newblock JHEP {\bf 01}, 068 (2006), arXiv:hep-ph/0512052.

\bibitem{Garbrecht:2014kda}
B.~Garbrecht and P.~Schwaller,
\newblock JCAP {\bf 10}, 012 (2014), arXiv:1404.2915.

\bibitem{Engelhard:2006yg}
G.~Engelhard, Y.~Grossman, E.~Nardi, and Y.~Nir,
\newblock Phys. Rev. Lett. {\bf 99}, 081802 (2007), arXiv:hep-ph/0612187.

\bibitem{Bertuzzo:2010et}
E.~Bertuzzo, P.~Di~Bari, and L.~Marzola,
\newblock Nucl. Phys. B {\bf 849}, 521 (2011), arXiv:1007.1641.

\bibitem{Domcke:2020quw}
V.~Domcke, K.~Kamada, K.~Mukaida, K.~Schmitz, and M.~Yamada,
\newblock Phys. Rev. Lett. {\bf 126}, 201802 (2021), arXiv:2011.09347.

\bibitem{Fong:2020fwk}
C.~S. Fong,
\newblock Phys. Rev. D {\bf 103}, L051705 (2021), arXiv:2012.03973.

\bibitem{Giudice:2003jh}
G.~F. Giudice, A.~Notari, M.~Raidal, A.~Riotto, and A.~Strumia,
\newblock Nucl. Phys. B {\bf 685}, 89 (2004), arXiv:hep-ph/0310123.

\bibitem{Gunji:2022xig}
Y.~Gunji, K.~Ishiwata, and T.~Yoshida,
\newblock JHEP {\bf 11}, 002 (2022), arXiv:2208.10086.

\bibitem{Murayama:1992ua}
H.~Murayama, H.~Suzuki, T.~Yanagida, and J.~Yokoyama,
\newblock Phys. Rev. Lett. {\bf 70}, 1912 (1993).

\bibitem{Murayama:1993xu}
H.~Murayama, H.~Suzuki, T.~Yanagida, and J.~Yokoyama,
\newblock Phys. Rev. D {\bf 50}, R2356 (1994), arXiv:hep-ph/9311326.

\bibitem{Murayama:1993em}
H.~Murayama and T.~Yanagida,
\newblock Phys. Lett. B {\bf 322}, 349 (1994), arXiv:hep-ph/9310297.

\bibitem{Hamaguchi:2001gw}
K.~Hamaguchi, H.~Murayama, and T.~Yanagida,
\newblock Phys. Rev. D {\bf 65}, 043512 (2002), arXiv:hep-ph/0109030.

\bibitem{Ellis:2003sq}
J.~R. Ellis, M.~Raidal, and T.~Yanagida,
\newblock Phys. Lett. B {\bf 581}, 9 (2004), arXiv:hep-ph/0303242.

\bibitem{Antusch:2004hd}
S.~Antusch, M.~Bastero-Gil, S.~F. King, and Q.~Shafi,
\newblock Phys. Rev. D {\bf 71}, 083519 (2005), arXiv:hep-ph/0411298.

\bibitem{Antusch:2009ty}
S.~Antusch, M.~Bastero-Gil, K.~Dutta, S.~F. King, and P.~M. Kostka,
\newblock Phys. Lett. B {\bf 679}, 428 (2009), arXiv:0905.0905.

\bibitem{Kadota:2005mt}
K.~Kadota and J.~Yokoyama,
\newblock Phys. Rev. D {\bf 73}, 043507 (2006), arXiv:hep-ph/0512221.

\bibitem{Nakayama:2013nya}
K.~Nakayama, F.~Takahashi, and T.~T. Yanagida,
\newblock Phys. Lett. B {\bf 730}, 24 (2014), arXiv:1311.4253.

\bibitem{Casas:2001sr}
J.~A. Casas and A.~Ibarra,
\newblock Nucl. Phys. B {\bf 618}, 171 (2001), arXiv:hep-ph/0103065.

\bibitem{Esteban:2020cvm}
I.~Esteban, M.~C. Gonzalez-Garcia, M.~Maltoni, T.~Schwetz, and A.~Zhou,
\newblock JHEP {\bf 09}, 178 (2020), arXiv:2007.14792.

\bibitem{Ibanez:1992aj}
L.~E. Ibanez and F.~Quevedo,
\newblock Phys. Lett. B {\bf 283}, 261 (1992), arXiv:hep-ph/9204205.

\bibitem{Fong:2010qh}
C.~S. Fong, M.~C. Gonzalez-Garcia, E.~Nardi, and J.~Racker,
\newblock JCAP {\bf 12}, 013 (2010), arXiv:1009.0003.

\bibitem{Fong:2012buy}
C.~S. Fong, E.~Nardi, and A.~Riotto,
\newblock Adv. High Energy Phys. {\bf 2012}, 158303 (2012), arXiv:1301.3062.

\bibitem{Fong:2015vna}
C.~S. Fong,
\newblock Phys. Lett. B {\bf 752}, 247 (2016), arXiv:1508.03648.

\bibitem{Domcke:2020kcp}
V.~Domcke, Y.~Ema, K.~Mukaida, and M.~Yamada,
\newblock JHEP {\bf 08}, 096 (2020), arXiv:2006.03148.

\bibitem{Akrami:2018odb}
Planck, Y.~Akrami {\em et~al.},
\newblock Astron. Astrophys. {\bf 641}, A10 (2020), arXiv:1807.06211.

\end{thebibliography}
\bibliographystyle{h-physrev5}

\end{document}